\documentclass[letterpaper]{JHEP3}

\usepackage{amsmath}

\title{Analytic Lifshitz black holes in higher dimensions}

\author{Eloy Ay\'on--Beato\\
Departamento~de~F\'{\i}sica,~CINVESTAV--IPN,\\
Apdo.~Postal~14--740,~07000,~M\'exico~D.F.,~M\'exico.\\
E-mail: \email{ayon-beato-at-fis.cinvestav.mx}}

\author{Alan Garbarz\\
Departamento de F\'{\i}sica, Universidad de Buenos Aires FCEN -
UBA,\\ Ciudad Universitaria, Pabell\'on 1, 1428,
Buenos Aires, Argentina. \\
E-mail: \email{alan-at-df.uba.ar}}

\author{Gaston Giribet\\
Departamento de F\'{\i}sica, Universidad de Buenos Aires FCEN -
UBA and CONICET,\\ Ciudad Universitaria, Pabell\'on 1, 1428,
Buenos Aires, Argentina. \\
E-mail: \email{gaston-at-df.uba.ar}}

\author{Mokhtar Hassa\"{\i}ne\\
Instituto de Matem\'atica y F\'{\i}sica, Universidad de Talca,\\
Casilla 747, Talca, Chile.\\
E-mail: \email{hassaine-at-inst-mat.utalca.cl}}

\abstract{We generalize the four-dimensional $R^2$--corrected
$z=3/2$ Lifshitz black hole to a two-parameter family of black
hole solutions for any dynamical exponent $z$ and for any
dimension $D$. For a particular relation between the
parameters, we find the first example of an extremal Lifshitz
black hole. An asymptotically Lifshitz black hole with a
logarithmic decay is also exhibited for a specific critical
exponent depending on the dimension. We extend this analysis to
the more general quadratic curvature corrections for which we
present three new families of higher-dimensional $D\geq 5$
analytic Lifshitz black holes for generic $z$. One of these
higher-dimensional families contains as critical limits the
$z=3$ three-dimensional Lifshitz black hole and a new $z=6$
four-dimensional black hole. The variety of
analytic solutions presented here encourages to
explore these gravity models within the context of
non-relativistic holographic correspondence.}

\preprint{\arXivid{1001.2361 [hep-th]}}

\begin{document}

\section{Introduction}

In the last years, promising attempts to extend the AdS/CFT
correspondence to other areas of physics have attracted much
attention. Within the context of non-relativistic physics,
holographic techniques were recently considered with remarkable
success. Pioneer work on this matter, where gravity duals to
non-relativistic systems were proposed, has been done in
Refs.~\cite{Son} and \cite{McGreevy}. Of particular importance
is also the construction of Ref.~\cite{Kachru}, where the
authors proposed new gravitational duals to scale invariant
Lifshitz fixed points with no Galilean invariance. These
gravity backgrounds have the form
\begin{equation}
ds^{2}=-\frac{r^{2z}}{l^{2z}}dt^{2}+\frac{l^{2}}{r^{2}}dr^{2}
+\frac{r^{2}}{l^{2}}d\vec{x}^2,  \label{V}
\end{equation}
where $\vec{x}$ is a $(D-2)$-dimensional vector. These
geometries are usually called Lifshitz spacetimes, and admit
the following anisotropic scaling symmetry
\begin{equation}\label{U}
t\mapsto\lambda^{z}t,\qquad r\mapsto\lambda^{-1}r,\qquad
\vec{x}\mapsto\lambda\vec{x},
\end{equation}
as part of their isometry group. This is the geometric
realization of the scale invariance exhibited by their
non-relativistic dual systems, which are thought to be
formulated on the $(D-1)$-dimensional space located at infinite
$r$. In this sense, this picture completely mimics the
prescription of the standard AdS/CFT correspondence.

The natural extension of the construction of \cite{Kachru} is
to look for black hole configurations that asymptote the
Lifshitz spacetimes (\ref{V}). Holographically, they should
describe the finite temperature behavior of the
non-relativistic theories. Black hole solutions of this type
are known with the name of ``Lifshitz black holes", and the
quest for such solutions has received much attention recently.
Analytic Lifshitz black hole solutions are scarce and they are
actually hard to be found. In spite of the fact their metrics
are not of a particularly abstruse form, these are reluctant to
appear as exact solutions of theories of gravity with
physically sensible matter sources. The main obstacle for these
spacetimes to exist are the Birkhoff theorems, which happen to
hold for generic models and restrict the subspace of static
solutions in a strong way. Nevertheless, some few examples of
black hole solutions that are asymptotically Lifshitz spaces
were recently found in the literature. One of the first
analytic examples was reported in Ref.~\cite{Taylor} for a sort
of higher-dimensional dilaton gravity without restricting the
value of the dynamical exponent $z$. In Ref.~\cite{Mann}, a
topological black hole solution which happens to be
asymptotically Lifshitz with $z=2$ was found. An example with
$z=4$ and with spherical topology was given in
Ref.~\cite{BertoldiI}. Numerical solutions for more general
values of $z$ were explored in
Refs.~\cite{DanielssonI,Mann,BertoldiI,BertoldiII}. More
examples of analytic Lifshitz black holes were studied in
Refs.~\cite{DanielssonII,Balasubramanian:2009rx,charged}, and
the solution found in \cite{Balasubramanian:2009rx} is
particularly interesting as it corresponds to a remarkably
simple analytic example with $z=2$ in $D=4$ dimensions. The
difficulty of embedding Lifshitz black holes in string theory
was also investigated in
Refs.~\cite{TakayanagiI,TakayanagiII,Tong}. The holographic
description of asymptotically Lifshitz spacetimes was studied
in \cite{Ross}. More recent investigations related to Lifshitz
black holes can be found in
Refs.~\cite{Sin:2009wi,elnuestro,Pang:2009wa,Otro,Otro2}.

In \cite{AyonBeato:2009nh} a remarkably simple solution with
$z=3$ in absence of matter fields was found for the New Massive
Gravity theory \cite{Bergshoeff:2009hq}, which consists of
special square-curvature corrections to three-dimensional
gravity. Previously, in Ref.~\cite{Adams:2008zk}, it was shown
that square-curvature corrections to gravity generically can
support the Lifshitz spacetimes (\ref{V}). The example of
Ref.~\cite{AyonBeato:2009nh} is the first to show that these
theories also allows the existence of Lifshitz black holes.
Another example with $z=3/2$ was subsequently found for a
four-dimensional theory with $R^2$--corrections in
Ref.~\cite{Cai:2009ac}.

Inspired in the results of \cite{AyonBeato:2009nh} and \cite{Cai:2009ac}, we will investigate in this paper how
the introduction of higher-curvature corrections to the Einstein-Hilbert action leads to find a large zoo of
analytic Lifshitz black hole solutions in $D$ dimensions. We will begin our search of Lifshitz black holes by
considering the simplest example of quadratic corrections. Already in the simplest case we will find interesting
solutions, provided a suitable parameterization of the coupling constants, and which hold for generic $z$ in $D$
dimensions. Interestingly enough, we will also exhibit an extremal Lifshitz black hole and an asymptotically
Lifshitz black hole with logarithmic decay at infinity. Motivated by the richness of examples we find in the
simplest case, we will then consider the most general square-curvature corrections, and we will present several
classes of analytic Lifshitz black hole families of solutions in $D\geq 5$ dimensions. Curiously, one of these
higher-dimensional families leads, through some particular limiting procedure, to the three-dimensional $z=3$
Lifshitz black hole of \cite{AyonBeato:2009nh} as well as to a new solution in $D=4$ with critical exponent $z=6$.

\section{$R^2$--corrected Lifshitz black holes for any dimension}

We first consider a gravity theory with $R^2$--corrections
\begin{equation}\label{eq:SquadR2}
S[g_{\mu\nu}]=\int{d}^Dx\sqrt{-g}
\left(R-2\lambda+\beta_1{R}^2\right),
\end{equation}
giving the following field equations
\begin{equation}\label{eq:eqR2}
G_{\mu\nu}+\lambda{g}_{\mu\nu}+2\beta_1g_{\mu\nu}\square{R}
-2\beta_1\nabla_\mu\nabla_\nu{R}+2\beta_1RR_{\mu\nu}
-\frac12\beta_1{R}^2g_{\mu\nu}=0.
\end{equation}

These equations allow Lifshitz spacetimes (\ref{V}) as
solutions for a generic value of the dynamical exponent $z$ in
any dimension, provided a suitable choice of the cosmological
constant $\lambda$ and the coupling constant $\beta_1$ that is
given by
\begin{subequations}\label{eq:LparamR2}
\begin{eqnarray}
\lambda &=& -\frac{2z^2+(D-2)(2z+D-1)}{4l^2},
\label{eq:Lparam_lambdaR2}\\
\beta_1 &=& -\frac{1}{8\lambda}.\label{eq:Lparam_R2}
\end{eqnarray}
\end{subequations}
It is well known that this kind of theory (\ref{eq:SquadR2})
can be generically mapped into scalar-tensor theories through a
conformal transformation of the metric with conformal factor
$\Omega^2=1+2\beta_1R$. However, for the particular choice of
the coupling constant (\ref{eq:LparamR2}), this trick does not
work since $R=4\lambda$ for Lifshitz spacetimes. In this sense,
the model corresponds to a genuine pure gravity theory. The
black hole solutions we will derive below present the same
degeneracy for the conformal transformation, and thus have no
scalar-tensor counterpart.

Our purpose is to explore whether there exist some black hole
solutions which asymptote the Lifshitz spacetimes (\ref{V}).
This analysis is motivated by the existence of a
four-dimensional Lifshitz black hole solution for these
theories with a specific value of the dynamical exponent
$z=3/2$ found in Ref.~\cite{Cai:2009ac}. In fact, we can show
that for any dimension $D$, there exists a two-parametric
family of solutions given by
\begin{subequations}\label{eq:F1}
\begin{eqnarray}
 ds^2&=&-\frac{r^{2z}}{l^{2z}}
\left(1-\frac{M^{-}l^{\alpha_-}}{r^{\alpha_-}}
+\frac{M^{+}l^{\alpha_+}}{r^{\alpha_+}}\right)dt^2
+\frac{l^2}{r^2}
\left(1-\frac{M^-l^{\alpha_{-}}}{r^{\alpha_{-}}}
+\frac{M^+l^{\alpha_{+}}}{r^{\alpha_{+}}}\right)^{-1}dr^2\nonumber\\
&&{}+\frac{r^2}{l^2}d\vec{x}^2,\label{eq:gF1}\\
&&\nonumber\\
\alpha_{\pm}&=&\frac{3z+2(D-2)\pm\sqrt{z^2+4(D-2)(z-1)}}2,
\label{eq:alphaF1}
\end{eqnarray}
\end{subequations}
for which the coupling constants are the same as in the purely
Lifshitz case (\ref{eq:LparamR2}). It is important to mention
that this family of geometries has the same constant scalar
curvature of the Lifshitz spacetimes.

First, it is clear from the expression of $\alpha_{\pm}$, given
by (\ref{eq:alphaF1}), that the dynamical exponent may take the
values $z\in(-\infty,z_{-}]\cup[z_{+},\infty)$ where
\begin{equation}\label{eq:zpm}
z_{\pm}=4-2D\pm 2\sqrt{(D-1)(D-2)}.
\end{equation}
On the other hand, the solution (\ref{eq:F1}) represents an
asymptotically  Lifshitz black hole for $\alpha_\pm>0$, and
this occurs for $z\geq z_+$. It is easy to see that the
four-dimensional solution of reference \cite{Cai:2009ac}
corresponds to the particular case $M^+=0$ with $z=3/2$, that
is $\lambda=-33/8l^2$ and $\alpha_-=3$.

For a specific relation between the constants $M^{\pm}$ given
by
\begin{equation}
M^{+}=\alpha_-(\alpha_+-\alpha_-)^{\frac{\alpha_+-\alpha_-}{\alpha_-}}
\left(\frac{M^-}{\alpha_+}\right)^{\frac{\alpha_+}{\alpha_-}},
\end{equation}
the solution (\ref{eq:F1}) has zero temperature, i.e. it is an
extremal black hole
\begin{eqnarray}
ds^2&=&-\frac{r^{2z}}{l^{2z}}\left[1-\frac{\alpha_+}{\alpha_+-\alpha_-}
\left(\frac{r_e}{r}\right)^{\alpha_-}+
\frac{\alpha_-}{\alpha_+-\alpha_-}
\left(\frac{r_e}{r}\right)^{\alpha_+}\right]dt^2\nonumber\\
&&{} +\frac{l^2}{r^2} \left[1-\frac{\alpha_+}{\alpha_+-\alpha_-}
\left(\frac{r_e}{r}\right)^{\alpha_-}+
\frac{\alpha_-}{\alpha_+-\alpha_-}
\left(\frac{r_e}{r}\right)^{\alpha_+}\right]^{-1}dr^2+
\frac{r^2}{l^2}d\vec{x}^2,\label{eq:extremal}
\end{eqnarray}
where the extremal radius $r_e$ is expressed as
\begin{equation}
r_e=l\left(\frac{\alpha_+-\alpha_-}{\alpha_+}M^-\right)^{1/\alpha_-}.
\end{equation}

The interest on solution (\ref{eq:F1}) increases once one
notices that when the dynamical exponent approach the value
$z=z_+$, defined in (\ref{eq:zpm}), there exists an additional
solution which asymptotes the Lifshitz spacetime in a much
slower way
\begin{eqnarray}
ds^2&=&-\frac{r^{2z_+}}{l^{2z_+}}
\left\{1-\frac{l^{\alpha_0}}{r^{\alpha_0}}
\left[M_1+M_2\ln{\left(\frac{r}{l}\right)}\right]\right\}dt^2
+\frac{l^2}{r^2}
\left\{1-\frac{l^{\alpha_0}}{r^{\alpha_0}}
\left[M_1+M_2\ln{\left(\frac{r}{l}\right)}\right]\right\}^{-1}dr^2
\nonumber\\
&&{}+\frac{r^2}{l^2}d\vec{x}^2,\label{eq:log}
\end{eqnarray}
where the parameter $\alpha_0$ is given by
\begin{equation}\label{eq:alpha0}
\alpha_0=3\sqrt{(D-1)(D-2)}-2(D-2).
\end{equation}
The fact of having a weakened (logarithmic) fall-off as a
next-to-leading contribution in the asymptotic behavior is well
known in the standard AdS/CFT correspondence. In particular,
this was one of the key points in recent discussions on
three-dimensional massive gravity (see \cite{Strominger} and
reference therein).

The extremal version of the logarithmic black hole
(\ref{eq:log}) is found for
$$
M_1=\frac{M_2}{\alpha_0}\left[1-\ln\left(\frac{M_2}{\alpha_0}\right)
\right],
$$
and then the corresponding spacetime geometry reads
\begin{eqnarray}
ds^2&=&-\frac{r^{2z_+}}{l^{2z_+}}
\left\{1-\frac{{r_e}^{\alpha_0}}{r^{\alpha_0}}
\left[1+\alpha_0\ln{\left(\frac{r}{r_e}\right)}\right]\right\}dt^2
+\frac{l^2}{r^2}
\left\{1-\frac{{r_e}^{\alpha_0}}{r^{\alpha_0}}
\left[1+\alpha_0\ln{\left(\frac{r}{r_e}\right)}\right]\right\}^{-1}
dr^2\nonumber\\
&&{}+\frac{r^2}{l^2}d\vec{x}^2,\label{eq:logext}
\end{eqnarray}
where the extremal radius is defined by
$$
r_e=l\left(\frac{M_2}{\alpha_0}\right)^{\frac{1}{\alpha_0}}.
$$
Let us stress that spacetimes (\ref{eq:extremal}) and (\ref{eq:logext}) are the first examples of asymptotically
Lifshitz black hole solutions with an extremal horizon. Remarkably, the solution (\ref{eq:log}) is additionally
the first solution with a logarithmic decay.

In the list of curiosities, we can also mention that the family
of solutions (\ref{eq:F1}) contains an asymptotically conformal
limit at $z=1$; namely
\begin{equation}\label{eq:STAdS}
ds^2=-\left(\frac{r^2}{l^2}
-\frac{M^-l^{D-2}}{r^{D-2}}+\frac{M^+l^{D-3}}{r^{D-3}}\right)dt^2
+\left(\frac{r^2}{l^2}-\frac{M^-l^{D-2}}{r^{D-2}}
+\frac{M^+l^{D-3}}{r^{D-3}}\right)^{-1}dr^2
+\frac{r^2}{l^2}d\vec{x}^2.
\end{equation}
For $M^-=0$ ($M^+<0$), the resulting spacetime is nothing but
the Schwarzschild-Tangherlini-AdS topological black hole with
toroidal horizon ($k=0$ in standard notation) for
$\lambda=-D(D-1)/(4l^2)$. For $M^+=0$, the solution corresponds
to a different asymptotically AdS toroidal black hole with a
faster decay. As an appealing remark, for $D=4$, the solution
(\ref{eq:STAdS}) is precisely the Reissner-Nordstrom-AdS
topological black hole with $k=0$.

As a final comment, we would like to point out that the
Lagrangian of the gravity action (\ref{eq:SquadR2}) for the
coupling constant given by (\ref{eq:Lparam_R2}) can be written
as a perfect square,
\begin{equation}
\label{eq:LF1}
R-2\lambda-\frac1{8\lambda}R^2=
-\frac{1}{8\lambda}\left(R-4\lambda\right)^2.
\end{equation}
Therefore, for this choice of the coupling constant, the
gravity action is definite positive and reaches its minimal
(vanishing) value for $R=4\lambda$, which is precisely the case
of the solutions (\ref{eq:F1}). Then, this solution (or its
Euclidean continuation) can be seen as a sort of gravitational
instanton. This has to do with the degeneracy of the field
equations (\ref{eq:eqR2}) in the following sense: for the case
of constant scalar curvature solutions, equations
(\ref{eq:eqR2}) become
$$
f^{\prime}(R)R_{\mu\nu}-\frac{1}{2}f(R)g_{\mu\nu}=0,
$$
where $f(R)$ is the Lagrangian expressed in terms of the scalar
curvature. Apart from considering $f^{\prime}(R)\not=0$, which
yields Einstein equations with an effective cosmological
constant and hence no Lifshitz configurations, the only option
for constant scalar curvature solutions to exist is that the
value of the scalar curvature be a double root of the
Lagrangian $f(R)$. Then, it is clear that the family  of black
holes obtained in (\ref{eq:F1}) will be solutions of any
gravity theory with Lagrangian $f(R)=(R-4\lambda)^2H(R)$ where
$H$ is a function regular at $R=4\lambda$.

\section{More general quadratic corrections}

Due to the new and interesting results of the $R^2$--corrected
theory (\ref{eq:SquadR2}) presented above, it is natural to
extend the analysis and explore the existence of asymptotically
Lifshitz black hole configurations with the most general
quadratic corrections. We will proceed in the same way as
before, by first establishing the purely Lifshitz
configurations and then by presenting three different classes
of asymptotically Lifshitz black hole solutions which
correspond to different ranges of the dynamical exponent $z$.
As shown below, for any value of the dynamical exponent $z$ at
least one family of black hole solutions exists.

We consider now the gravity action that includes the most
general quadratic-curvature corrections in $D$-dimensions;
namely
\begin{eqnarray}
S[g_{\mu\nu}]&=&\int{d}^Dx\sqrt{-g} \left(R-2\lambda+\beta_1{R}^2
+\beta_2{R}_{\alpha\beta}{R}^{\alpha\beta}
+\beta_3{R}_{\alpha\beta\mu\nu}{R}^{\alpha\beta\mu\nu} \right).
\label{eq:Squad}
\end{eqnarray}
It follows from the Gauss-Bonnet theorem in four dimensions and
the vanishing of the Gauss-Bonnet term in three dimensions that
in the case $D < 5$, it is sufficient to consider only two of
the three quadratic invariants in the Lagrangian. This makes
necessary to split the analysis in two parts by first
considering the higher-dimensional cases and then to analyze
separately the lower dimensional ones, $D=3$ and $D=4$.

The action (\ref{eq:Squad}) gives rise to the following field
equations
\begin{eqnarray}
G_{\mu\nu}+\lambda{g}_{\mu\nu}
+\left(\beta_2+4\beta_3\right)\square{R}_{\mu\nu}
+\frac12\left(4\beta_1+\beta_2\right)g_{\mu\nu}\square{R}
-\left(2\beta_1+\beta_2+2\beta_3\right)\nabla_\mu\nabla_\nu{R}
\nonumber\\ \nonumber\\
{}+2\beta_3R_{\mu\gamma\alpha\beta}R_{\nu}^{~\gamma\alpha\beta}
+2\left(\beta_2+2\beta_3\right)R_{\mu\alpha\nu\beta}R^{\alpha\beta}
{}-4\beta_3R_{\mu\alpha}R_{\nu}^{~\alpha}+2\beta_1RR_{\mu\nu}
\nonumber\\ \nonumber\\
{}-\frac12\left(\beta_1{R}^2+\beta_2{R}_{\alpha\beta}{R}^{\alpha\beta}
+\beta_3{R}_{\alpha\beta\gamma\delta}{R}^{\alpha\beta\gamma\delta}
\right)g_{\mu\nu}&=&0.\qquad \label{eq:squareGrav}
\end{eqnarray}
As in the purely $R^2-$case, Lifshitz spacetimes (\ref{V}) are
solutions of these field equations for a generic value of the
dynamical exponent $z$ in any dimension, provided that
\begin{subequations}\label{eq:Lparam}
\begin{eqnarray}
\lambda &=&
-\frac1{4l^2}\bigg(2z^2+(D-2)(2z+D-1)-\frac{4(D-3)(D-4)z(z+D-2)\beta_3}{l^2}\bigg),
\label{eq:Lparam_lambda}\\\nonumber\\
\beta_2 &=& \frac{l^2-2\left[2z^2+(D-2)(2z+D-1)\right]\beta_1
-4\left[z^2-(D-2)z+1\right]\beta_3}
{2(z^2+D-2)}.\label{eq:Lparam_beta}
\end{eqnarray}
\end{subequations}
Notice that the above parameterizations coincide with the
values previously found for the New Massive Gravity in $D=3$
\cite{AyonBeato:2009nh}, where $\beta_3=0$ and
$\beta_2=-(8/3)\beta_1=-1/m^2$. In turn, this generalizes the
previous authors' result. We shall now proceed to present three
more different Lifshitz black hole families in $D\geq5$
dimensions.

\subsection{An asymptotically Lifshitz black hole family for $z>2-D$}

The first family of solutions we present here is described by
the following line element
\begin{subequations}\label{eq:F2}
\begin{equation}\label{eq:gF2}
ds^2=-\frac{r^{2z}}{l^{2z}}
\left(1-\frac{Ml^{(z+D-2)/2}}{r^{(z+D-2)/2}}\right)dt^2+\frac{l^2}{r^2}
\left(1-\frac{Ml^{(z+D-2)/2}}{r^{(z+D-2)/2}}\right)^{-1}dr^2
+\frac{r^2}{l^2}d\vec{x}^2,
\end{equation}
and represents an asymptotically Lifshitz black hole solution of the field equations (\ref{eq:squareGrav}) for the
dynamical exponent $z>2-D$. The coupling constants allowing the existence of the solution (\ref{eq:gF2}) are
parameterized in term of the dynamical exponent $z$ by
\begin{eqnarray}
\lambda&=&\frac{(D-2)}{4l^2}{\Big\{}(197D-389)z^4
+4(19D^2-200D+325)z^3+(D-2)\big[2(5D^2-73D+356)z^2
\nonumber\\\nonumber\\
&&{}+4(D^3-2D^2+15D-62)z+(D+2)(D-1)(D-2)^2\big]{\Big\}} \bigg/
P_4(z),\label{eq:lambdaF2}\\\nonumber\\
\beta_1&=&l^2{\Big\{}27z^6-18(3D-4)z^5
+3(19D^2-168D+356)z^4-12(11D^3-84D^2+196D-120)z^3
\nonumber\\\nonumber\\
&&{}-(D-2)\big[(19D^3-330D^2+2052D-3640)z^2
+2(3D^4-30D^3+124D^2-536D+1024)z\nonumber\\\nonumber\\
&&{}+(D+2)(D-2)^2(D^2-4D+36) \big]{\Big\}}
\bigg/{\Big(2(D-3)(D-4)(z+D-2)^2 P_4(z)\Big)},
\label{eq:beta1F2}\\\nonumber\\
\beta_2&=&-2l^2\big[3z^2+(D+2)(D-2)\big]
{\Big\{}9z^4-6(3D-4)z^3-8(D^2-10)z^2+2(D^3-4D^2+32D-80)z
\nonumber\\\nonumber\\
&&{}-(D-2)\big[D^3+2D^2-12(D-2)\big]{\Big\}}
\bigg/{\Big((D-3)(D-4)(z+D-2)^2 P_4(z)\Big)},
\label{eq:beta2F2}\\\nonumber\\
\beta_3&=&l^2\big[3z^2+(D+2)(D-2)\big] {\Big\{}9z^3-3(9D-14)z^2
-(D-2)\big[(5D-62)z+D^2-4D+36\big]{\Big\}}
\nonumber\\\nonumber\\
&&{}\bigg/{\Big(2(D-3)(D-4)(z+D-2) P_4(z)\Big)}, \label{eq:beta3F2}
\end{eqnarray}
\end{subequations}
where $P_4$ is a polynomial of degree four in the dynamical
exponent $z$ given by
$$
P_4(z)=27z^4-4(27D-45)z^3
-(D-2)\big[2(5D-116)z^2+4(D^2-D+30)z+(D+2)(D-2)^2\big].
$$
It is clear from the expressions of the $\beta_i$ that the
solution is defined only for higher dimensions $D\ge5$. The
analysis of the lower dimensional cases will be done in the
next section. As in the purely $R^2$--case, there exists a
conformal limit $z=1$ of the family (\ref{eq:F2}) which is
given by the following asymptotically AdS black hole
\begin{equation}\label{eq:gz=1F2}
ds^2=-\left(\frac{r^2}{l^2}
-\frac{Ml^{(D-5)/2}}{r^{(D-5)/2}}\right)dt^2+\left(\frac{r^2}{l^2}
-\frac{Ml^{(D-5)/2}}{r^{(D-5)/2}}\right)^{-1}dr^2
+\frac{r^2}{l^2}d\vec{x}^2.
\end{equation}
More precisely, this spacetime  (\ref{eq:gz=1F2}) is a solution
of the Einstein-Gauss-Bonnet gravity with a fine-tuned coupling
constant, since for $z=1$ the parameterizations
(\ref{eq:lambdaF2})-(\ref{eq:beta3F2}) become
\begin{subequations}
\begin{eqnarray}
\lambda&=&-\frac{(D-1)(D-2)}{4l^2},\label{eq:lambdaz=1F2}\\
\beta_1=-\frac14\beta_2=\beta_3&=&\frac{l^2}{2(D-3)(D-4)},
\label{eq:betasz=1F2}
\end{eqnarray}
\end{subequations}
yielding to the Lagrangian
\begin{equation}\label{eq:L_iCS}
R-2\lambda -\frac{(D-1)(D-2)}{8(D-3)(D-4)\lambda}{\cal
L}_{\mathrm{GB}},
\end{equation}
where ${\cal L}_{\mathrm{GB}}=R^2
-4R_{\alpha\beta}{R}^{\alpha\beta}
+{R}_{\alpha\beta\mu\nu}{R}^{\alpha\beta\mu\nu}$ is the
Gauss-Bonnet term. Note that for $D=5$, the above black hole
becomes diffeomorphic to a warped product having as base
AdS$_3$ with a 2-plane fiber. Moreover this case precisely
corresponds to the Chern-Simons gravity in $D=5$.

\subsection{An asymptotically Lifshitz black hole family for $z>1$}

The second family of asymptotically Lifshitz black holes we
find is valid for $z>1$,
\begin{subequations}\label{eq:F3}
\begin{eqnarray}
ds^2&=&-\frac{r^{2z}}{l^{2z}}
\left(1-\frac{Ml^{2(z-1)}}{r^{2(z-1)}}\right)dt^2+\frac{l^2}{r^2}
\left(1-\frac{Ml^{2(z-1)}}{r^{2(z-1)}}\right)^{-1}dr^2
+\frac{r^2}{l^2}d\vec{x}^2,\label{eq:gF3}
\end{eqnarray}
and it exists for the following choice of coupling constants
\begin{eqnarray}
\lambda&=&-\frac{(z-1)\left[z^2-Dz-(D-1)(D-2)\right]}{2l^2(z-D)},
\label{eq:lambdaF3}\\\nonumber\\
\beta_1&=&l^2\Big[3(D-1)(D-2)z^3
-(2D^3-2D^2-11D+20)z^2+(3D^3-14D^2+19D+10)z
\nonumber\\\nonumber\\
&&{}+(D+2)(D-4)\Big]\Big/{\Big[}2(D-2)(D-3)(D-4)z(z-1)(z-D)(3z+D-4)
{\Big]},
\label{eq:beta1F3}\\\nonumber\\
\beta_2&=&-l^2(D-1)(2z-D-2) {\Big[}6(D-2)z^2-(D^2-3D+8)z-2(D-4)
{\Big]}\nonumber\\\nonumber\\
&&{}\Big/{\Big[}2(D-2)(D-3)(D-4)z(z-1)(z-D)(3z+D-4) {\Big]},
\label{eq:beta2F3}\\\nonumber\\
\beta_3&=&\frac{l^2(D-1)(2z-D-2)}{4(D-3)(D-4)z(z-D)}.
\label{eq:beta3F3}
\end{eqnarray}
\end{subequations}
As for the $z>2-D$ family, the lower dimensions $D=3$ and $D=4$
are also forbidden here. Clearly, there is no conformal limit
$z=1$ for this family.

\subsection{An asymptotically Lifshitz black hole family for  $z<0$}

The last family of Lifshitz black holes we describe is
characterized by a negative dynamical critical exponent
$z=-|z|$, whose metric reads
\begin{subequations}\label{eq:F4}
\begin{eqnarray}
ds^2&=&-\frac{l^{2|z|}}{r^{2|z|}}
\left(1-\frac{Ml^{|z|}}{r^{|z|}}\right)dt^2
+\frac{l^2}{r^2} \left(1-\frac{Ml^{|z|}}{r^{|z|}}\right)^{-1}dr^2
+\frac{r^2}{l^2}d\vec{x}^2,\label{eq:gF4}
\end{eqnarray}
while the corresponding coupling constants are parameterized as
\begin{eqnarray}
\lambda&=&\frac{|z|\left[2|z|^2-4(D-2)|z|+(D-2)(D-3)\right]}
               {4l^2(2|z|-D+2)},\qquad~
\label{eq:lambdaF4}\\\nonumber\\
\beta_1&=&\frac{l^2\left[3|z|^2-2(D-2)|z|+2D-5\right]}
               {2(D-3)(D-4)|z|(2|z|-D+2)},
               \label{eq:beta1F4}\\\nonumber\\
\beta_2&=&-4\beta_3,\\\nonumber\\
\beta_3&=&\frac{l^2\left[6|z|^2-4(D-2)|z|+(D-1)(D-2)\right]}
               {4(D-3)(D-4)|z|(2|z|-D+2)}.\qquad~
\label{eq:beta23F4}
\end{eqnarray}
\end{subequations}
The associated Lagrangian describes a fine-tuned
$R^2$--corrected Einstein-Gauss-Bonnet theory
\begin{equation}\label{eq:L_z<0}
R-2\lambda-\beta_3{\cal L}_{\mathrm{GB}}
-\frac{l^2R^2}{4|z|(2|z|-D+2)}.
\end{equation}
This family of black holes is again defined only in higher
dimensions $D\ge5$. Being defined only for negative dynamical
critical exponents, it has no conformal analog $z=1$.

In the next section, we analyze the lower-dimensional cases
$D=3$ and $D=4$.

\section{Critical lower dimensional Lifshitz black holes}

The families of Lifshitz black holes given by (\ref{eq:F2}),
(\ref{eq:F3}) and (\ref{eq:F4}) are generically forbidden in
dimensions lower than $5$. This is due to the fact that the use
of a nontrivial value for the coupling constant $\beta_3$ is
artificial in these dimensions. Concretely, if one consider
theories with $\beta_3\neq0$, and due to the fact that the
Gauss-Bonnet combination ${\cal L}_{\mathrm{GB}}$ vanishes in
$D=3$ and is a total derivative in $D=4$, it turns out that it
is always possible to shift the coupling constants and to end
with $\beta_3=0$. This shifting reads
\begin{equation}\label{eq:beta_shift}
(\beta_1,\beta_2,\beta_3)\mapsto(\beta_1-\beta_3,\beta_2+4\beta_3,0).
\end{equation}

Despite families (\ref{eq:F2}), (\ref{eq:F3}) and (\ref{eq:F4}) are formally defined for higher-dimensions, the
possibility that new critical solutions exist in lower dimensions for some particular values of the dynamical
exponent $z$ is not excluded. A natural way to explore this possibility is to consider a dimensional continuation
of the $D$-dimensional expressions and study whether some potential cancellation of the divergences of the
coupling constants appears when one expands around $D=3$ and $D=4$. That is what we will do in this section. Using
the results of this analysis as an indication, we will explicitly confirm the existence of critical solutions
that, indeed, represent Lifshitz black holes in $D=3$ and in $D=4$.

\subsection{The $z=3$ three-dimensional asymptotically Lifshitz
black hole}

Let us start with the dimensional continuation and expansion of
the coupling constants of the family (\ref{eq:F2}) around
$D=3$. The cosmological constant is regular,
$\lambda=O\left((D-3)^0\right)$, but the coupling constants
exhibit the following singular behavior
\begin{eqnarray}\label{eq:betasF2D=3}
\beta_1=-\frac14\beta_2=\beta_3&=&
-\frac{(z-3)(3z^2+5)(9z^2-12z+11)l^2}
{2(z+1)(27z^4-144z^3+202z^2-144z-5)}\times\frac1{D-3}\nonumber\\
\nonumber\\
&&{}+O\left((D-3)^0\right).
\end{eqnarray}
This indicates that the only possibility for having a
potentially regular behavior for this family at $D=3$ appears
for $z=3$. Considering that there is in fact no continuity in
the number of dimensions, one can chose the element $z=3$ of the family
(\ref{eq:F2}). Evaluating after that for $D=3$, the resulting
Lifshitz black hole is
\begin{subequations}\label{eq:NMG}
\begin{equation}\label{eq:gNMG}
ds^2=-\frac{r^{6}}{l^{6}}
\left(1-\frac{Ml^{2}}{r^{2}}\right)dt^2+\frac{l^2}{r^2}
\left(1-\frac{Ml^{2}}{r^{2}}\right)^{-1}dr^2
+\frac{r^2}{l^2}d{x}^2,
\end{equation}
with $\lambda=-13/(2l^2)$, and surprisingly the meaningful
coupling constants (i.e.\ after the shifting
(\ref{eq:beta_shift})) are those of New Massive Gravity
\cite{Bergshoeff:2009hq}
\begin{equation}\label{eq:bNMG}
\beta_2=-(8/3)\beta_1=2l^2,
\end{equation}
\end{subequations}
which gives rise to the three-dimensional Lifshitz black hole
previously found by the authors in \cite{AyonBeato:2009nh}. A
similar analysis can be done for the family (\ref{eq:F3}); the
potential regular behavior occurs in this case for $z=5/2$.
However, the resulting solution is not new but corresponds to
the case $z=5/2$, $M^+=0$ ($\alpha_-=3$) of the family
(\ref{eq:F1}) valid in generic dimension $D$. The family
(\ref{eq:F4}) has no regular limit in $D=3$.

\subsection{The $z=6$ four-dimensional asymptotically Lifshitz
black hole}

In four dimensions, we proceed in a similar way, by doing a
dimensional continuation and expanding the coupling constants
of the family (\ref{eq:F2}) around $D=4$. Again, the
cosmological constant is regular,
$\lambda=O\left((D-4)^0\right)$, and the coupling constants
exhibit singular behavior
\begin{eqnarray}\label{eq:betasF2D=4}
\beta_1=-\frac14\beta_2=\beta_3&=&
\frac{3(z-6)(z^2+4)(3z^2-4z+4)l^2}
{2(z+2)(9z^4-84z^3+128z^2-112z-16)}\times\frac1{D-4}\nonumber\\
\nonumber\\
&&{}+O\left((D-4)^0\right).
\end{eqnarray}
The indication here is that the only possibility potentially occurs for $z=6$. The $z=6$ element of the family
(\ref{eq:F2}), when is evaluated in $D=4$, indeed gives rise to a new Lifshitz black hole
\begin{subequations}\label{eq:LbhD=4}
\begin{equation}\label{eq:gLbhD=4}
ds^2=-\frac{r^{12}}{l^{12}}
\left(1-\frac{Ml^{4}}{r^{4}}\right)dt^2+\frac{l^2}{r^2}
\left(1-\frac{Ml^{4}}{r^{4}}\right)^{-1}dr^2
+\frac{r^2}{l^2}(d{x}^2+d{y}^2),
\end{equation}
with $\lambda=-51/(2l^2)$ and meaningful coupling constants given by
\begin{equation}\label{eq:bLbhD=4}
\beta_2=-(25/9)\beta_1=25l^2/64.
\end{equation}
\end{subequations}
The corresponding analysis for the family (\ref{eq:F3}) in
$D=4$ singles out the value $z=3$. Again, the resulting
solution is not new but corresponds to the case $z=3$, $M^+=0$
($\alpha_-=4$) of the family (\ref{eq:F1}), which is valid in
four dimensions. As before, family (\ref{eq:F4}) has no regular
limit in $D=4$.

\section{Conclusions and open problems}

In this paper we have found analytic Lifshitz black hole
solutions for gravity with square-curvature corrections in
arbitrary dimension. Some open questions remain:

\begin{itemize}
\item The computation of conserved charges of the
    asymptotically Lifshitz black holes of higher-curvature
    gravity would be needed to fully understand the
    thermodynamical properties of both the gravitational
    backgrounds and the dual systems. Some important
    advances in this direction have  been done recently in
    \cite{Hohm}.

\item Stability of Lifshitz black hole solutions is another
    question it would be interesting to address.

\item Among the family of black holes we exhibited here
    there are extremal solutions, see (\ref{eq:extremal})
    and (\ref{eq:logext}). An interesting question is that
    of studying the causal structure of these spacetimes.

\item Last, the condensed matter interpretation of these
    backgrounds within the holographic proposal of
    \cite{Kachru} deserves to be matter for further study.
\end{itemize}

\begin{acknowledgments}
The authors thank the organizers and participants of the CECS
Summer Workshop on Theoretical Physics. The work of A.G. is
supported by University of Buenos Aires, Argentina. G.G. is
Member of the CIC-CONICET, Argentina. This work has been
partially supported by grant 1090368 from FONDECYT, by grants
UBACyT X861 X432 from UBA, by grant PICT 00849 from ANPCyT, and
by grants 82443 and 45946-F from CONACyT.
\end{acknowledgments}


\begin{thebibliography}{99}

\bibitem{Son} D.T. Son, \textit{Toward an AdS/cold atoms
    correspondence: a geometric realization of the Schroedinger
    symmetry}, Phys. Rev. \textbf{D78} (2008) 046003,
    [arXiv:0804.3972].

\bibitem{McGreevy} K. Balasubramanian and J. McGreevy,
    \textit{Gravity duals for non-relativistic CFTs}, Phys.
    Rev. Lett. \textbf{101} (2008) 061601, [arXiv:0804.4053].

\bibitem{Kachru} S. Kachru, X. Liu and M. Mulligan,
    \textit{Gravity Duals of Lifshitz-like Fixed Points}, Phys.
    Rev. \textbf{D78} (2008) 106005, [arXiv:0808.1725].

\bibitem{Taylor} M. Taylor, \textit{Non-relativistic
    holography}, [arXiv:0812.0530].

\bibitem{Mann} R. Mann, \textit{Lifshitz Topological Black
    Holes}, [arXiv:0905.1136].

\bibitem{BertoldiI} G. Bertoldi, B. Burrington and A. Peet,
    \textit{Black Holes in asymptotically Lifshitz spacetimes
with arbitrary critical exponent}, [arXiv:0905.3183].

\bibitem{DanielssonI} U. Danielsson and L. Thorlacius,
    \textit{\ Black holes in asymptotically Lifshitz
    spacetime}, JHEP \textbf{0903} (2009) 070,
    [arXiv:0812.5088].

\bibitem{BertoldiII} G. Bertoldi, B. Burrington and A. Peet,
    \textit{Thermodynamics of black branes in asymptotically
    Lifshitz spacetimes}, [arXiv:0907.4755].

\bibitem{DanielssonII} E. Brynjolfsson, U. Danielsson, L.
    Thorlacius and T. Zingg, \textit{Holographic
    Superconductors with Lifshitz Scaling}, [arXiv:0908.2611].

\bibitem{Balasubramanian:2009rx}
  K.~Balasubramanian and J.~McGreevy,
  {\it An analytic Lifshitz black hole},
  [arXiv:0909.0263].

\bibitem{charged} Da-Wei Pang, \textit{On Charged Lifshitz
    Black Holes}, [arXiv:0911.2777].

\bibitem{TakayanagiI} T. Azeyanagi, W. Li and T. Takayanagi,
    \textit{On String Theory Duals of Lifshitz-like Fixed
    Points}, JHEP \textbf{0906} (2009) 084, [arXiv:0905.0688].

\bibitem{TakayanagiII} W. Li, T. Nishioka and T. Takayanagi,
    \textit{Some No-go Theorems for String Duals of
    Non-relativistic Lifshitz-like Theories},
    [arXiv:0908.0363].

\bibitem{Tong} S. Hartnoll, J. Polchinski, E. Silverstein and
    D. Tong, {\it Towards strange metallic holography},
    [arXiv:0912.1061].

\bibitem{Ross} S. Ross and O. Saremi {\it Holographic stress
    tensor for non-relativistic theories} JHEP {\bf 0909}
    (2009) 009, [arXiv:0907.1846].

\bibitem{Sin:2009wi}
  S.~J.~Sin, S.~S.~Xu and Y.~Zhou,
  \textit{Holographic Superconductor for a Lifshitz fixed
  point}, 
  arXiv:0909.4857 [hep-th].

\bibitem{elnuestro} Y. Myung, Y-W. Kim and Y-J. Park,
    \textit{Dilaton gravity approach to three dimensional
    Lifshitz black hole}, [arXiv:0910.4428].

\bibitem{Pang:2009wa}
  D.~W.~Pang, \textit{Conductivity and Diffusion Constant
  in Lifshitz Backgrounds},
  arXiv:0912.2403 [hep-th].

\bibitem{Otro} M. Cheng, S. Hartnoll and C. Keelerar,
    \textit{Deformations of Lifshitz holography},
    [aXiv:0912.2784].




\bibitem{Otro2} K.B. Fadafan, \textit{Drag force in
    asymptotically Lifshitz spacetimes}, [arXiv:0912.4873].




\bibitem{AyonBeato:2009nh}
  E.~Ay\'{o}n-Beato, A.~Garbarz, G.~Giribet and M.~Hassa\"{\i}ne,
  {\it Lifshitz Black Hole in Three Dimensions},
  Phys. Rev. \textbf{D80} (2009) 104029,
  [arXiv:0909.1347].

\bibitem{Bergshoeff:2009hq}
  E.~A.~Bergshoeff, O.~Hohm and P.~K.~Townsend,
  {\it Massive Gravity in Three Dimensions},
  Phys. Rev. Lett. {\bf 102} (2009) 201301,
  [arXiv:0901.1766].

\bibitem{Adams:2008zk}
  A.~Adams, A.~Maloney, A.~Sinha and S.~E.~V\'{a}zquez,
  {\it 1/N Effects in Non-Relativistic Gauge-Gravity Duality},
  JHEP {\bf 0903} (2009) 097, [arXiv:0812.0166].

\bibitem{Cai:2009ac}
  R.~G.~Cai, Y.~Liu and Y.~W.~Sun,
  {\it A Lifshitz Black Hole in Four Dimensional $R^2$ Gravity},
  JHEP \textbf{0910} (2009) 080,
  [arXiv:0909.2807].

\bibitem{Strominger} A. Maloney, W. Song and A. Strominger,
    {\it Chiral Gravity, Log Gravity and Extremal CFT},
    [arXiv:0903.4573].

\bibitem{Hohm} O. Hohm and E. Tonni, \textit{A boundary stress tensor for
higher-derivative gravity in AdS and Lifshitz backgrounds}, to appear.

\end{thebibliography}
\end{document}